\documentclass{article}
\usepackage[dvips]{graphicx,psfrag}

\begin{document}
\title{
  Exact Phase Solutions  of
  Nonlinear Oscillators on Two-dimensional Lattice
  }
\author{
  Tsunehiro Yokoi, Hiroyasu Yamada, and Kazuhiro Nozaki
  }
\date{}
\maketitle

\begin{abstract}
  We present various exact solutions of a discrete
  complex Ginzburg-Landau (CGL) equation on a 
plane lattice, which describe target patterns 
and spiral patterns and derive their stability criteria.
We also obtain similar  solutions to a system of 
van der Pol's oscillators.
\end{abstract}

\section{Introduction}
Near the onset of the Hopf bifurcation in continuous media, a complex 
amplitude of
distributed oscillators is described by a normal form equation called  
the complex Ginzburg-Landau (CGL) equation. Some exact solutions of the 1+1 
dimensional
CGL equation are known  as  hole, solitary-wave and shock solutions 
\cite{nozaki-bekki}. In the 2+1 dimensional space, numerical studies 
showed existence of interesting solutions describing target patterns 
and spiral patterns  \cite{kramer1,kramer2}.
However,  explicit
and  exact solutions representing target patterns or spiral patterns 
have not  been constructed yet.

  Here  we consider a system of  oscillators which undergo the Hopf 
bifurcation on a two dimensional lattice instead  of continuous media. 
Each oscillator is governed by the normal form equation of the Hopf 
bifurcation \cite{gholms} and couples diffusively and dispersively with 
its nearest-neighbor  oscillators so that the system becomes a 
spatially discretized  version of the 2+1 dimensional CGL (discrete 
CGL) equation\cite{mori}.
Such a discrete  CGL equation has also been studied on 
a network in connection with \cite{strogatz,yamada}. In this paper,
  we show that the discrete  CGL equation has exact solutions 
describing various
  phase patterns consisting of target patterns and spiral patterns. Each 
spiral pattern solution needs  a defect in its origin of spiral, where 
  oscillators are absent.
  This defect may corresponds to the singularity of  the origin in a 
spiral solution
  in  continuous media. In the other hand, target pattern solutions do 
not have any defects.
  We derive  the stability condition of the exact solutions to 
modulational perturbations
  and conduct some numerical simulations to confirm the stability 
criterion. Similar phase pattern solutions are constructed to a system 
of van der Pol's oscillators.

 %%%%%%%
\section{Nonlinear Oscillators on Two-dimensional Lattice }
%%%%%%%%
Let us consider the following system of normal form oscillators to the 
Hopf bifurcation
on a two-dimensional lattice.
\begin{equation}
  \frac{dW_n}{dt} = -(\mu_r +i \mu_i)|W_n|^2 W_n
   +(\nu_r+ i \nu_i)W_n
   +(\lambda_r +i \lambda_i)\sum_{j \sim  n} ^4 (W_j - W_n) \label{eq:1}
  \end{equation}
where
$W_n $ are complex amplitudes of oscillators and
$ \mu _r,\ \mu_i,\ \nu_r,\ \nu_i,\ \lambda_r,\ \lambda _i $
are real constants; $ \mu _r$ 
and $ \nu _r$ are assumed to be positive (the  subcritical bifurcation) and
% $W_n \in {\it C}, \mu _r,\mu _i,\nu_r,\nu_i,\lambda_r ,\lambda _i \in {\it R}$
  $\sum^4 _{j \sim n}$ denotes summation over nearest-neighbor 
oscillators
  at site $n$. Through this summation term, each oscillator has 
diffusive ($\lambda _r$) and dispersive  ($\lambda _i$) interactions 
with its nearest-neighbors. Here, we call
   Eq.(\ref{eq:1})
  a discrete CGL equation.

By changing variables
$W_n  \rightarrow W_n \exp[-i \nu _i t]$ and $t \rightarrow \mu_r t$,
and replacing coefficients as
$p_1 = \mu _i /\mu_r $, $ p_2 = \nu_r/\mu_r  $,
$p_3 = \lambda_r/ \mu_r$ and $p_4 = \lambda_i/\mu_r$,
Eq.(\ref{eq:1}) is transformed into
\begin{equation}
  \frac{dW_n}{dt} = -(1 +i p_1)|W_n|^2 W_n
   +p_2 W_n+(p_3 +i p_4)\sum^4_{j \sim  n}(W_j - W_n) ,\label{eq:2}
\end{equation}
or
\begin{equation}
  \frac{dW_n}{dt} = -(1 +i p_1)|W_n|^2 W_n
   +(p_2 -4(p_3 +i p_4))W_n+(p_3 +i p_4)\sum^4_{j \sim  n}W_j 
,\label{eq:3}
\end{equation}
where $p_2>0$.
We use Eq.(\ref{eq:2}) or Eq.(\ref{eq:3}) as a basic system in this 
paper.

%%%%%%%%%
\section{Exact Solutions}
%%%%%%%%%%
In this section, we derive some exact solutions of the discrete CGL 
equation (\ref{eq:3}),
which represent a target pattern, a spiral pattern and various mixtures 
of those patterns.
A crucial step to find exact solutions is to assume
\begin{equation}
   \sum^4_{j \sim n} W_j =0
   \ \mbox{(for all sites}\  n \mbox{)}.\label{eq:4}
\end{equation}
Then, Eq. (\ref{eq:3}) takes the same equation for all $n$:
\begin{equation}
   \frac{dW_n}{dt}=-(1 +i p_1)|W_n|^2 W_n +(p_2-4 (p_3 +i p_4)) W_n 
.\label{eq:5}
\end{equation}
We choose an equilibrium  solution of Eq.(\ref {eq:5}).
\begin{equation}
   W_n =A e^{i(\omega t +\phi(n))} ,\label{eq:6}
\end{equation}
where $A ,\ \omega $ are real constants satisfying
\begin{eqnarray}
  A^2 &=& p_2-4p_3 , \\
  \omega &=& -p_1 (p_2 -4 p_3)-4 p_4 ,
\end{eqnarray}
and a functional form of the phase $\phi(n)$ with respect to $n$ is 
determined later.
Because of $A^2 >0$,  we need
\begin{equation}
p_2-4p_3>0. \label{cond1}
\end{equation}
Since $p_2>0$, the condition (\ref {cond1}) is automatically satisfied 
for
$p_3 \le 0$ (negative diffusion or diffusionless case).
The solution (\ref {eq:6}) has the same amplitude $A$ at  all cites but
its phase could vary from site to site. The variation of the phase must 
be determined
so that Eq.(\ref {eq:4}) is satisfied.
\begin{equation}
   \sum^4_{j \sim n} e^{i \phi(j)}=0
   \ \mbox{(for all sites}\  n \mbox{)}.\label{eq:7}
\end{equation}
We consider a case that there are only two values for $\phi (n)$,
that is, $\phi(n)=0$ or $\pi$. At every site, we set phase values of two oscillators  
  among four nearest-neighbor oscillators as $\phi(n)=0$, while the 
other two phases are chosen to be $\pi$.
Then, the condition (\ref {eq:7}) is fulfilled at all sites, that is,
\begin {eqnarray}
   \sum^4_{j \sim n} e^{i \phi(j)} &=&
   e^{i \cdot 0} +e^{i \cdot 0} +e^{i \cdot \pi} +e^{i \cdot 
\pi}\nonumber \\
  &=&0 \ \mbox{(for all sites}\  n \mbox{)} .\label {exact1}
\end{eqnarray}
This choice of phase values is quite simple but produces a variety of 
phase patterns
such as  stripe patterns and target patterns as exact solutions.
We can construct not only a single target pattern but also a combined 
pattern of
some target patterns as shown in Fig.1. Although we  present only a 
combined pattern of
four target patterns,  combined patterns of more or less target 
patterns are also
possible. We can construct  various combined patterns of  target 
patterns  indefinitely if the system size becomes larger.

When the lattice has a defect, that is, an oscillator is absent at a 
site, we should  modify the above discussion  for four oscillators 
around a defect.
To four oscillators around a defect, the coupling terms in 
Eq.(\ref{eq:2}) are changed to
\begin{eqnarray}
   (p_3 +i p_4) \sum_{j \sim  n} ^3 ( W_j - W_n )
   &=& (p_3 +i p_4)( \sum_{j \sim n} ^3 W_j -3 W_n) \nonumber \\
   &=& (p_3 +i p_4)\left\{ (W_n + \sum_{j \sim n} ^3 W_j)-4 W_n 
\right\},
\end{eqnarray}
  where $n$ indicates one of the four oscillators and the summation term 
in Eq.(\ref{eq:3}) becomes
\begin{equation}
   (p_3 +i p_4)(W_n + \sum_{j \sim n} ^3 W_j )  .
\end{equation}
Thus, Eq.(\ref{eq:7})  is modified for each $n$-th oscillator  around a 
defect as
\begin{equation}
     e^{i \phi(n)}+\sum_{j \sim n} ^3 e^{i \phi(j)}=0 .\label{eq:7m}
\end{equation}
Comparing Eq.(\ref{eq:7m}) to Eq.(\ref{eq:7}), the term 
$\exp (i \phi(n))$  in (\ref{eq:7m})
appears replacing the term which  comes from an oscillator at a defect 
site.
Therefore, for the $n$-th oscillator around a defect, the defect 
behaves as a mirror,
that is, the defect acts as if it is  an oscillator with the same 
phase as the $n$-th phase.
Owing to this flexible property of a defect, we can construct spiral 
patterns with $2^k$ arms around $k \times k$ combined defects. For 
example,  we obtain a spiral pattern with two arms around a single 
defect ($k=1$) and a spiral pattern with four arms around four combined 
defects ($k=2$) as shown in Fig.2 and Fig.4, respectively. Two separated defects produce
two spiral patterns with the same or opposite rotation directions as 
shown in Fig.3.
We can also construct various  patterns in which spirals and targets 
are concurrent and show one of the simplest examples  in Fig.2. On a 
large enough lattice, we could construct virtually an indefinite number 
of such spiral and target patterns.

\begin{figure}[htbp]
  \begin{tabular}{cc}
   \begin{minipage}{0.5\hsize}
    \begin{center}
     \includegraphics[width=3.5cm]{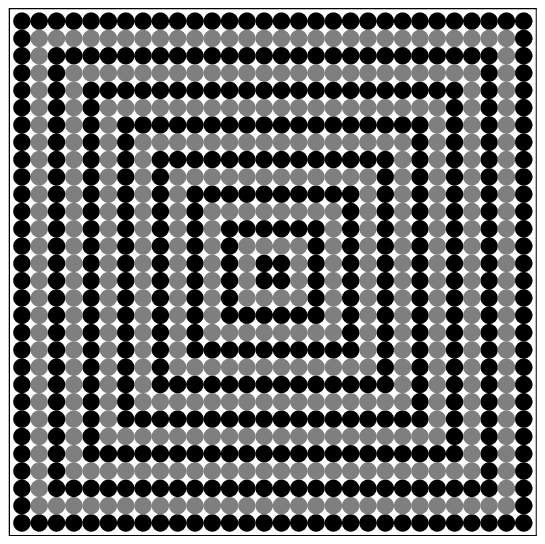}

    \end{center}
   \end{minipage}
   \begin{minipage}{0.5\hsize}
    \begin{center}
     \includegraphics[width=3.5cm]{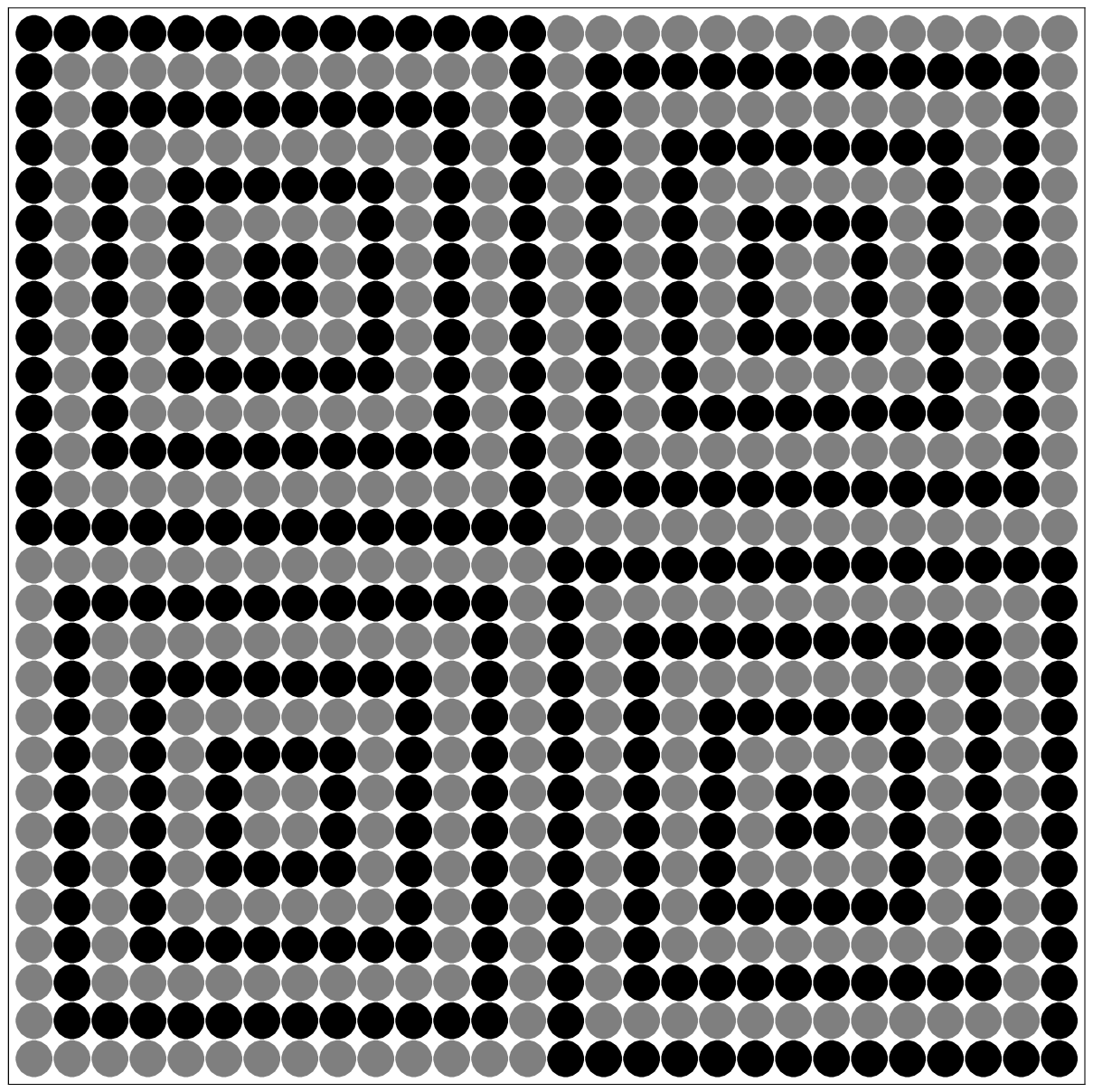}
   \end{center}
  \end{minipage}
\end{tabular}
\caption{A target pattern.
   Black points denote $\phi (n) =0$, while
   gray points denote $\phi (n) =\pi$.}
\label{fig:1}
\end{figure}

\begin{figure}[htbp]
  \begin{minipage}{0.5\hsize}
    \begin{center}
     \includegraphics[width=3.5cm]{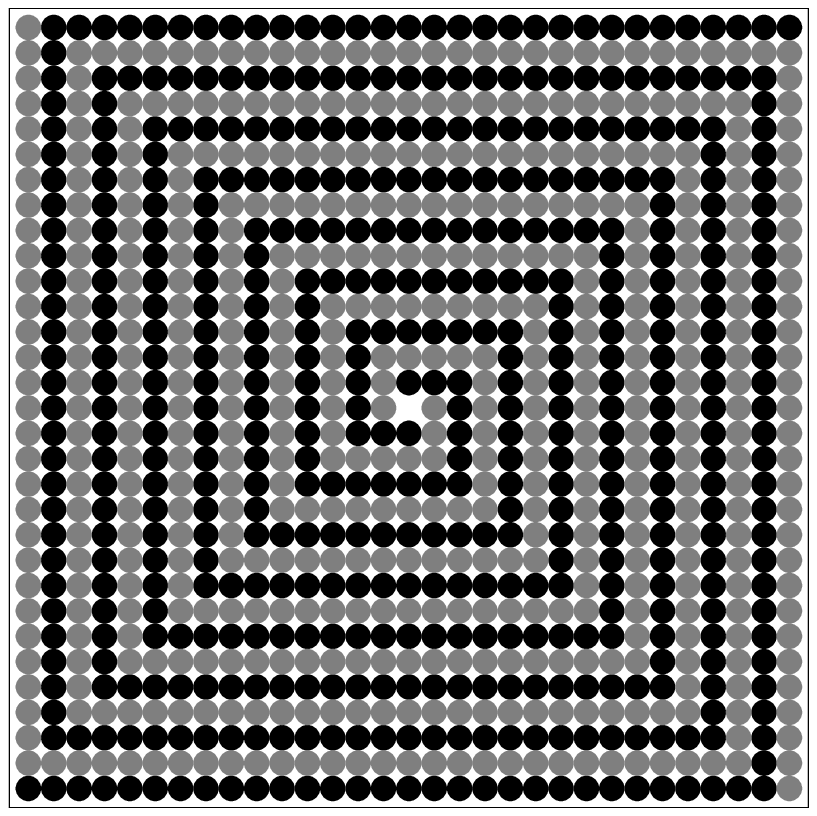}

    \end{center}
  \end{minipage}
   \begin{minipage}{0.5\hsize}
    \begin{center}
     \includegraphics[width=3.5cm]{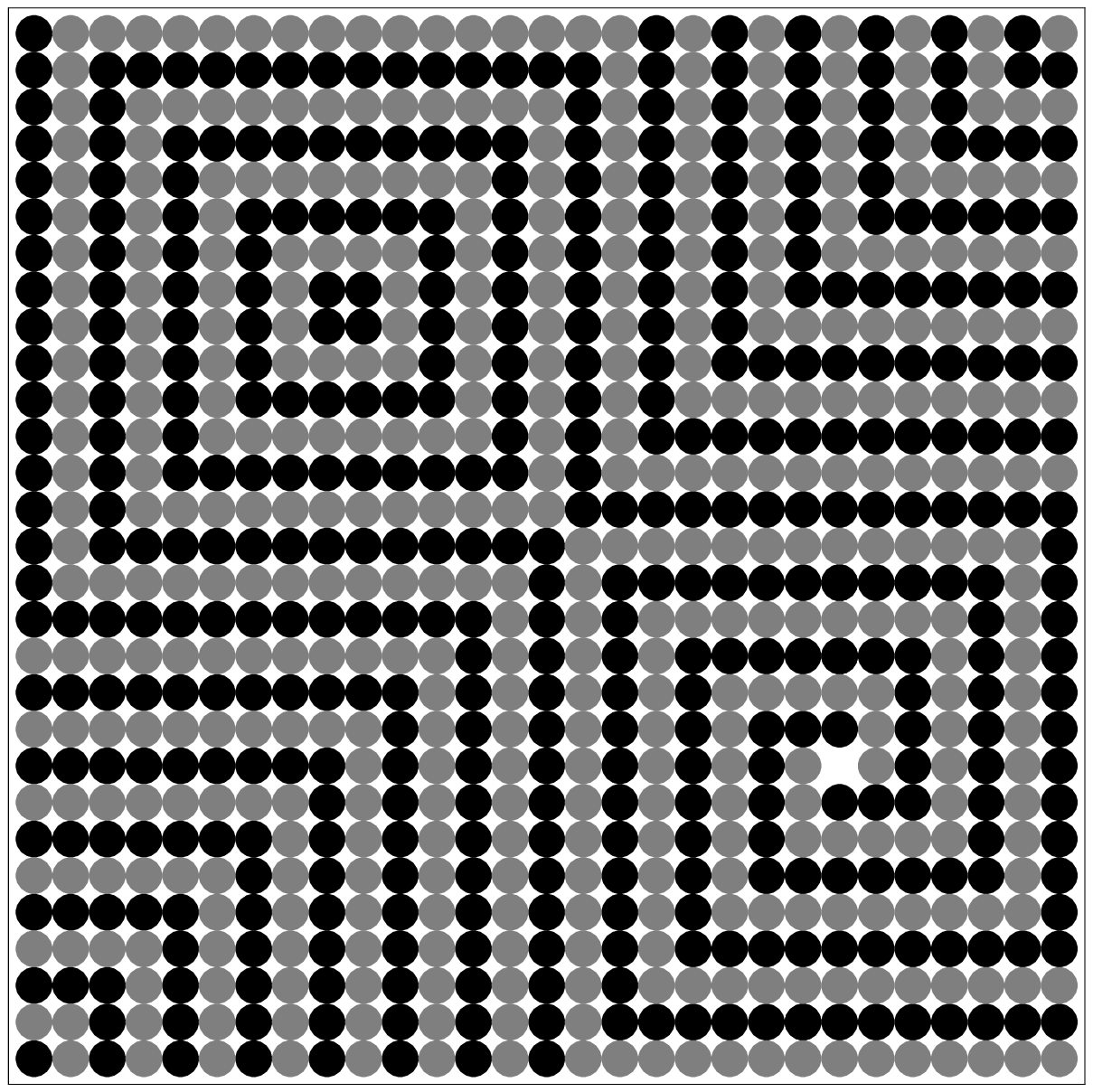}

    \end{center}
  \end{minipage}
      \caption{A spiral pattern with a defect and
      a combined pattern of a spiral and a target.}
     \label{fig:2}
\end{figure}

\begin{figure}[htbp]
  \begin{tabular}{cc}
   \begin{minipage}{0.5\hsize}
    \begin{center}
     \includegraphics[width=3.5cm]{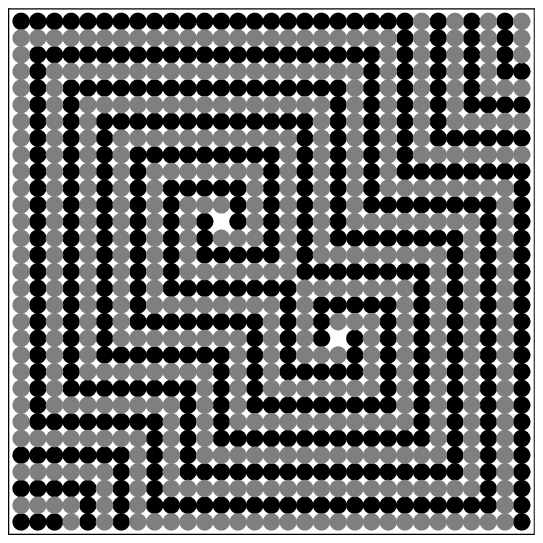}
    \end{center}
   \end{minipage}
   \begin{minipage}{0.5\hsize}
    \begin{center}
     \includegraphics[width=3.5cm]{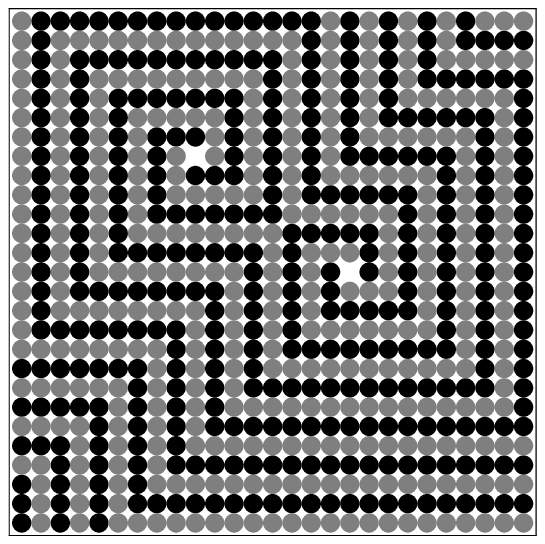}
    \end{center}
   \end{minipage}
\end{tabular}
     \caption{ Two spiral patterns  with  the opposite and the same
     spiral rotations, respectively.}
     \label{fig:3}
   \end{figure}
   
\begin{figure}[htbp]
  \begin{tabular}{cc}
   \begin{minipage}{0.5\hsize}
    \begin{center}
     \includegraphics[width=3.5cm]{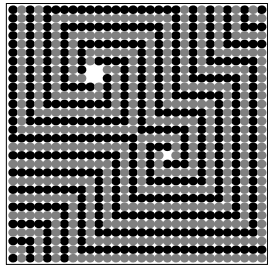}
    \end{center}
     \caption{A four-arm spiral pattern  and a two-arm spiral pattern}
   \label{fig:3.5}
  \end{minipage}
\end{tabular}

\end{figure}

%%%%%%%%%
\section{Stability of Solutions}
\begin{figure}[htbp]
  \begin{tabular}{cc}
   \begin{minipage}{0.45\hsize}
    \begin{center}
     \psfrag{a}[][][0.6]{$ \alpha _+$}
     \psfrag{K}[][][0.6]{K}
     \includegraphics[width=5.0cm]{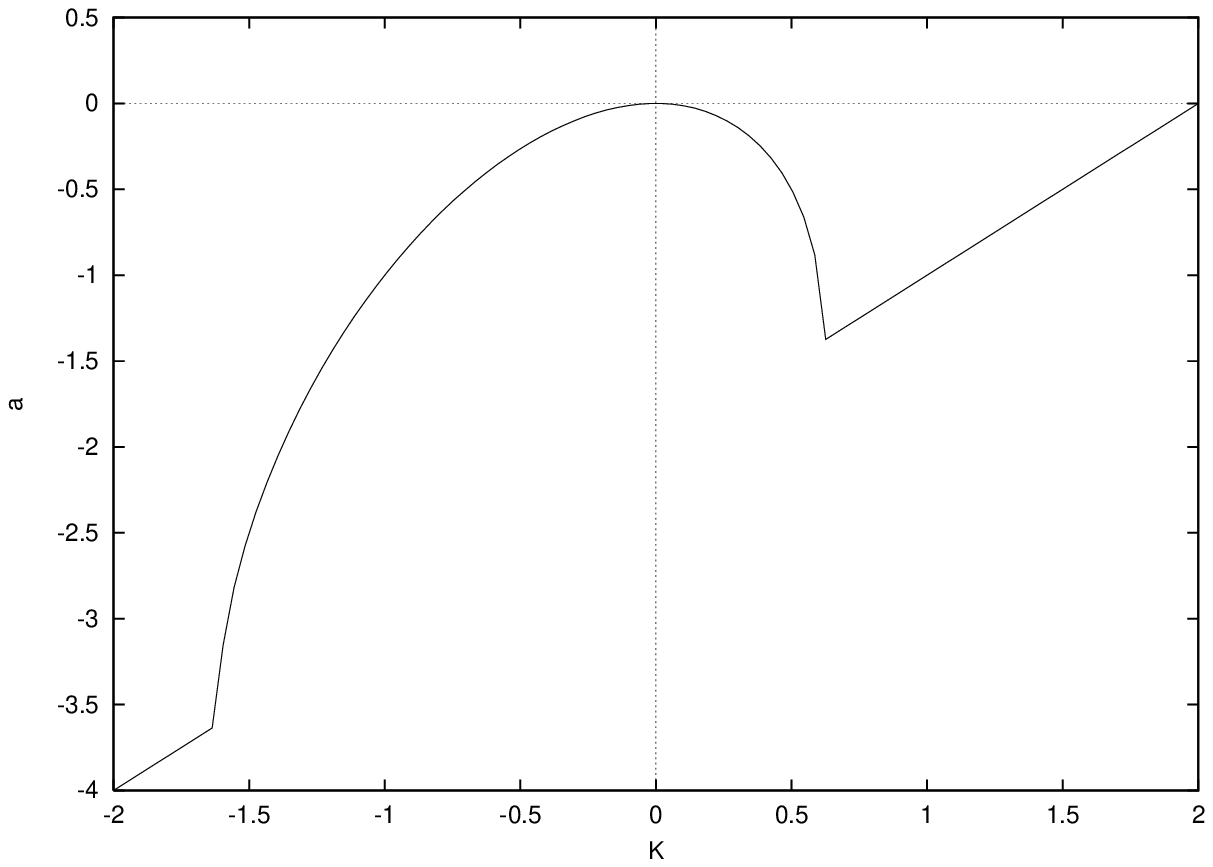}
     \caption{$\alpha_+ (K)$ vs. $K$ for stable parameters
       ($p_1 =-0.5$, $p_2 =4.0$, $p_3=0.5$, $p_4 =1.0$).}
     \label{fig:5}
    \end{center}
  \end{minipage}
  \begin{minipage}{0.1\hsize}
    \end{minipage}
   \begin{minipage}{0.45\hsize}
    \begin{center}
     \psfrag{a}[][][0.6]{$ \alpha _+$}
     \psfrag{K}[][][0.6]{K}
     \includegraphics[width=5cm]{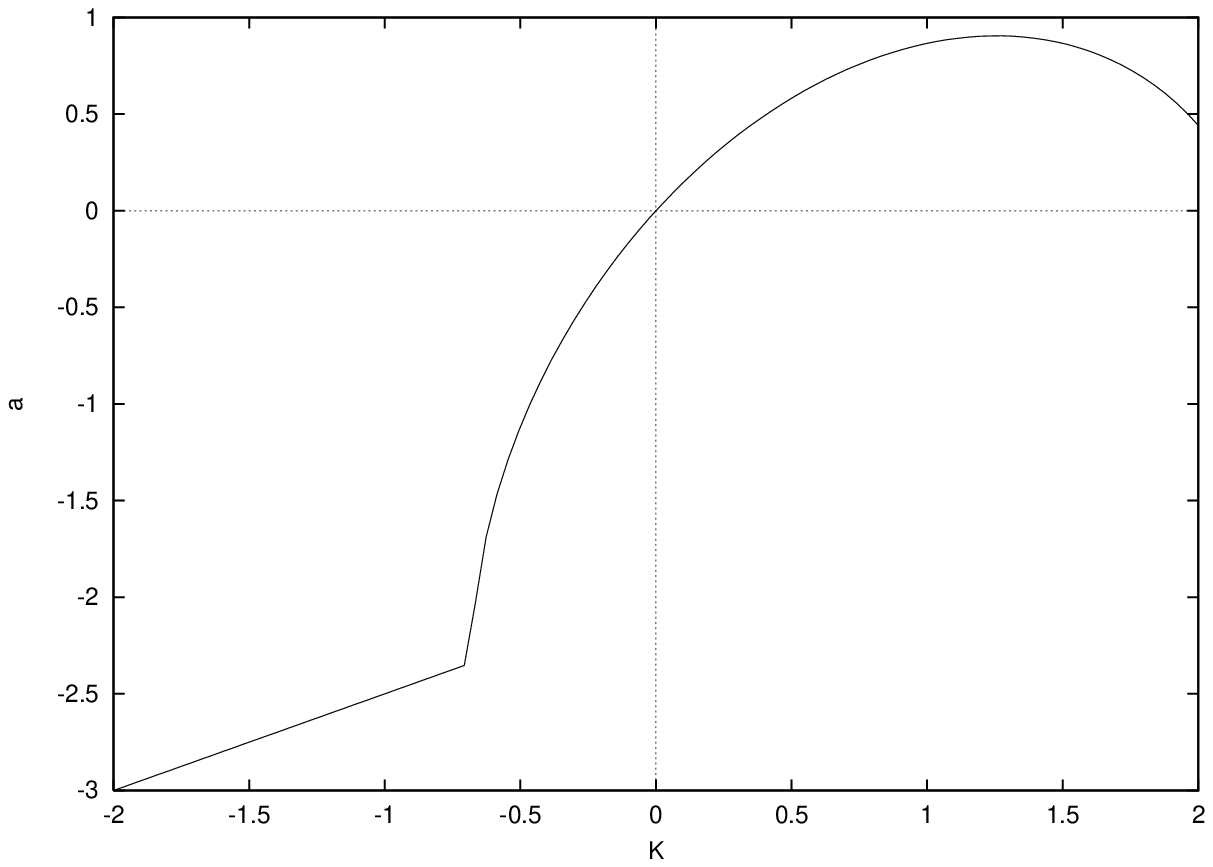}
     \caption{$\alpha_+ (K)$ vs. $K$  for unstable parameters
     ($p_1 =1.3$, $p_2 =3.0$, $p_3=0.25$, $p_4 =0.8$).}
     \label{fig:6}
    \end{center}
   \end{minipage}
  \end{tabular}
\end{figure}

%%%%%%%%%%
Let us  investigate a linear stability of the exact solution:
\begin{eqnarray}
&&  W_n ^0 (t) =
   A e^{ i(\omega t +\phi(n))}\ (\phi(n)=0\ \mbox{or}\  \pi), \nonumber\\
   &&  \sum^4_{j \sim n} e^{i \phi(j)}=0
   \ \mbox{(for all site}\  n \mbox{)}.\nonumber
\end{eqnarray}
Substituting a perturbed solution $W_n =W_n ^0 +\delta W_n$
into Eq.(\ref{eq:3}), we have
\begin{eqnarray}
  \frac{d}{dt}\delta W_n&=& i \omega \delta W_n-(1+i p_1)A^2 \delta W_n 
\nonumber\\
&&  -(1+i p_1)A^2 e^{2 i(\omega t+\phi(n))} \delta W_n ^*
   +(p_3 +i p_4) \sum^4_{j \sim n}\delta W_j. \label{eq:8}
\end{eqnarray}
As $\phi (n)=0$ or $\pi$, the  coefficient of the third term in the
RHS of Eq.(\ref {eq:8}) becomes
$-(1+i p_1)A^2 \exp[i 2 \omega t]$, which does not depend on
a site number $n$. Therefore, all coefficients of the linearized 
equation
(\ref {eq:8}) are independent of $n$, which greatly simplifies  
 stability analysis.
Changing the variable as
$\delta W_n \rightarrow \delta W_n \exp[-i \omega t]$,
Eq.(\ref{eq:8}) is transformed into
\begin{equation}
  \frac{d}{dt}\delta W_n= -(1+i p_1)A^2(\delta W_n + \delta W_n ^*)
   +(p_3 +i p_4) \sum^4_{j \sim n}\delta W_j. \label{eq:9}
\end{equation}
We consider plane-wave perturbations
\begin{equation}
   \delta W_n=a(t)e^{i \vec{k} \cdot \vec{x}}
   =a(t) e^{i (k_x x+k_y y)} ,\label{eq:10}
\end{equation}
where $\vec{x}=(x,y)$ indicates a lattice point and $\vec{k}=(k_x,k_y)$ 
is a wave-number vector.
Then, the summation term of Eq.(\ref{eq:9}) reads
\begin{eqnarray}
  \sum^4_{j \sim n} \delta W_j &=&
   a e^{\vec{k} \cdot \vec{x}}
   (e^{i k_x}+e^{i k_y}+e^{-i k_x}+e^{-i k_y}) \nonumber \\
  &=& a e^{\vec{k} \cdot \vec{x}}(2 \cos k_x + 2 \cos k_y) \nonumber \\
  &=&2 \delta W_n K, \nonumber
\end{eqnarray}
  where $K = \cos k_x +\cos k_y$.
Hence, Eq.(\ref{eq:9})  reduces to
\begin{eqnarray}
  \frac{d}{dt}\delta W_n  &=& -(1+i p_1)A^2(\delta W_n + \delta W_n ^*)
   +2 (p_3 +i p_4) K \delta W_n \nonumber \\
  &=&  (2 (p_3 +i p_4) K-(1+i p_1)A^2)\delta W_n
   -(1+i p_1)A^2 \delta W_n ^* ,
   \end{eqnarray}
or  in terms of the real and imaginary parts of $\delta W_n$
\begin{eqnarray}
  \frac{d}{dt} \left(
                \begin{array}{c} \Re [\delta W_n]\\
                \Im [\delta W_n]\end{array}
               \right)=2
  \left(
   \begin{array}{cc} p_3 K-A^2 & - p_4 K  \\
    p_4 K -p_1 A^2 & p_3 K\end{array}
  \right)
  \left(
   \begin{array}{c} \Re [\delta W_n]\\
   \Im [\delta W_n]\end{array}
  \right).\label{eq:11}
\end{eqnarray}
Eigenvalues of Eq.(\ref {eq:11}) are
\begin{equation}
  2 p_3 K-A^2 \pm \sqrt{-4  p_4 ^2 K^2
   +4 p_1 p_4 A^2 K+A^4}.
\end{equation}
We denote an eigenvalue as $\alpha_+$ which has a larger real part
than the other.
\begin{equation}
  \alpha_+ =2 p_3 K-A^2 +
   \Re \left[p
       \sqrt{-4  p_4 ^2 K^2
       +4 p_1  p_4 A^2 K+A^4}
      \right]. \label{eq:14}
\end{equation}
If $\alpha_+ \leq 0$ for any $K$ in $|K|\leq 2$,
the various phase solutions  are stable to a perturbation (\ref{eq:10}).
For stability, $\alpha_+$ must take  local maximum at K=0 since 
  $\alpha_+ =0$ at $K=0$
(see Fig.\ref{fig:5}, Fig.\ref{fig:6}).
The derivative of $\alpha_+$ at $K=0$ is given as
\begin{eqnarray}
\left.\frac{\partial \alpha_+}{\partial K} \right| _{K=0}
&=& 2 \left[
  p_3  -p_4 \frac{2 p_4 K- p_1 A^2}{
              \sqrt{-4 p_4 ^2 K^2 +4 p_1 p_4 K+A^4 }
              }
         \right]_{K=0} \nonumber \\
  &=& 2 (p_3+ p_1 p_4). \nonumber
\end{eqnarray}
A condition  $p_3+p_1 p_4=0$ is necessary for stability. 
When $p_3+p_1 p_4=0$, we have
\begin{eqnarray}
\left.\alpha _+ \right| _{K= \frac{A^2}{2 p_3}}
&=& \Re \left[
  \sqrt{- \left(\frac{p_4}{p_3} \right)^2
    + 2 \frac{p_1 p_4}{p_3} +1} \right] A^2\nonumber \\
  &=& \Re \left[ \sqrt{- \left(\frac{p_4}{p_3} \right)^2 -1} \right] A^2 \nonumber\\
  &=&0,\nonumber
\end{eqnarray}
which shows that  $K=A^2 /2 p_3 $ is also a vanishing point of $\alpha _+$, where its derivative with respect to $K$ is finite.
For stability, we need $$ |A^2 /2 p_3|\geq 2$$ or 
\begin{equation}
A^2=p_2 -4 p_3  \geq 4 |p_3|,  \label{cond-sta}
\end{equation}
Since $p_2>0$ is supposed, the condition (\ref {cond-sta})
reads
$ p_3 <0$  or $p_2>8p_3>0$.

Thus, we obtain
the following conditions for our  solutions to be stable.
\begin{eqnarray}
&&  p_3+p_1 p_4=0 \label{eq:15} \\ 
&&p_3 <0\  \mbox{or}\  p_2\geq 8p_3>0.\nonumber
 % A^2=p_2 -4 p_3 & \geq & 4 |p_3| \nonumber
\end{eqnarray}

%%%%%%%%%%%%%%%%%%%%%%%
\section{Numerical analysis of the system}
\begin{figure}[htbp]
   \begin{minipage}{0.5\hsize}
     \begin{center}
       \includegraphics[width=5.0cm]{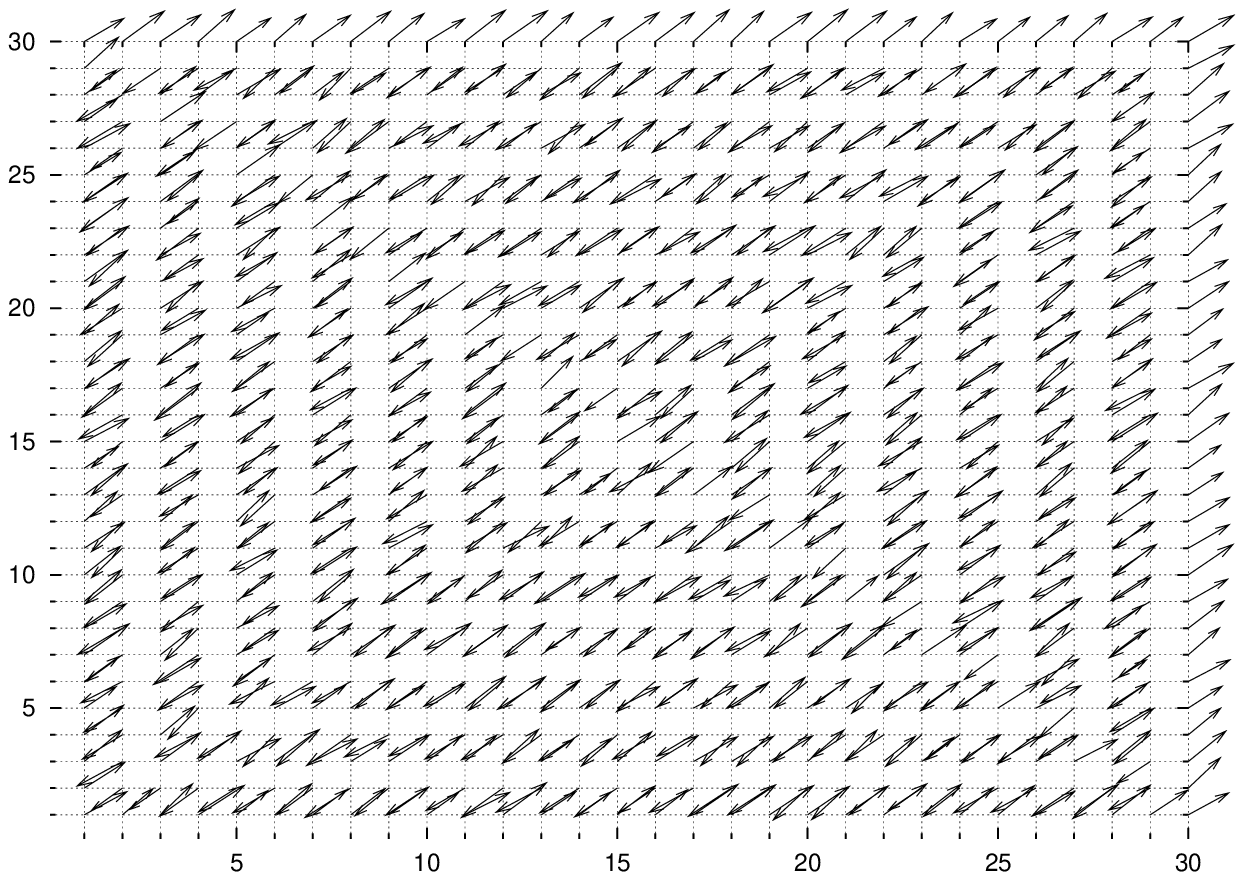}
     \end{center}
   \end{minipage}
     \begin{minipage}{0.5\hsize}
       \begin{center}
         \includegraphics[width=5.0cm]{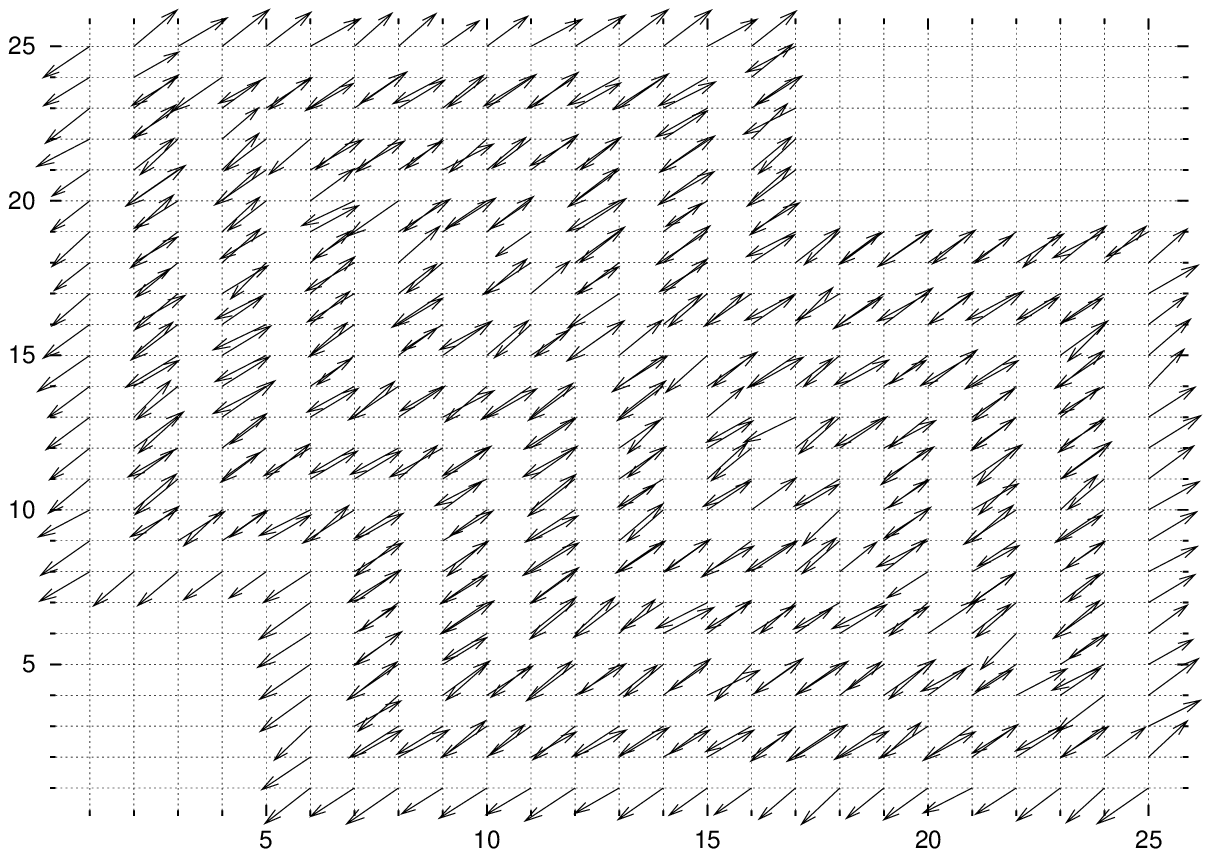}
       \end{center}
     \end{minipage}
     \caption{Initial 20\% random perturbations
       are added to  exact solutions in a
         stable case that  $p_1 =-0.5,p_2 =4.0,p_3=0.5,p_4 =1.0$.
         Length of an arrow indicates $|W_n |$
         and its direction  means $\arg W_n$ .}
       \label{fig:7}
\end{figure}

%%%%%%%%%%
First, we we conduct numerical analysis of the system with random 
initial perturbations in order to confirm validity of the stability 
condition (\ref{eq:15}).
  We set a boundary condition which enables
  our exact solutions  to persist. It is supposed
  that, outside a boundary,  there are oscillators whose phases are
different from those just inside  the boundary by $\pi$. This boundary 
condition is called a vanishing boundary condition in this paper.
For values of parameters satisfying the stability condition 
(\ref{eq:15}),
both target patterns and spiral patterns are found to be
stable even for   random perturbations.
For example, in a stable case that  $p_1 =-0.5$, $p_2 =4.0$, $p_3=0.5$ 
and $p_4 =1.0$, a single target pattern and the pattern with two 
spirals are  stable even for max $20 \%$ random perturbations
as shown in Fig.\ref{fig:7}.

\begin{figure}[htbp]
   \begin{tabular}{cc}
     \begin{minipage}{0.5\hsize}
       \begin{center}
         \includegraphics[width=5.0cm]{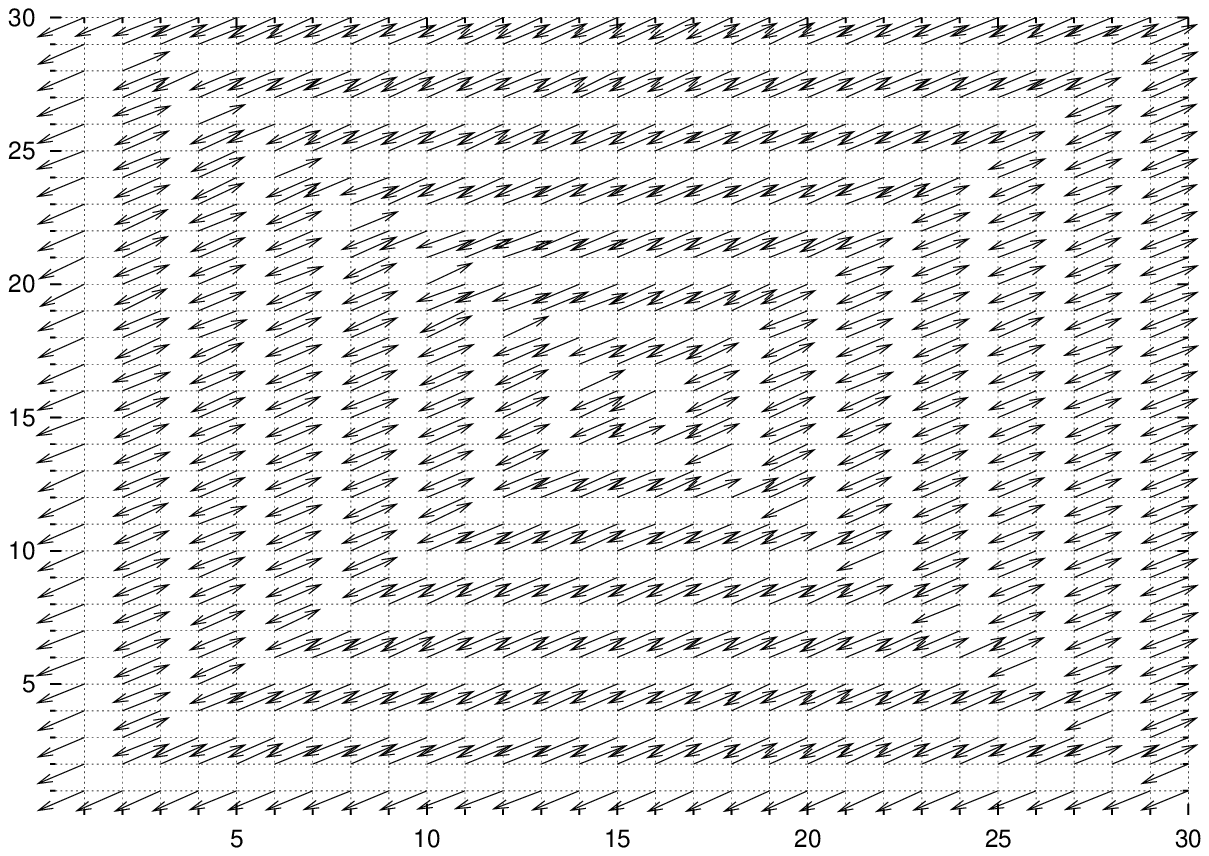}
       \end{center}
     \end{minipage}
     \begin{minipage}{0.5\hsize}
       \begin{center}
         \psfrag{amp}[][][0.6]{$|W|^2$}
         \psfrag{time}[][][0.6]{time}
         \includegraphics[width=5.0cm]{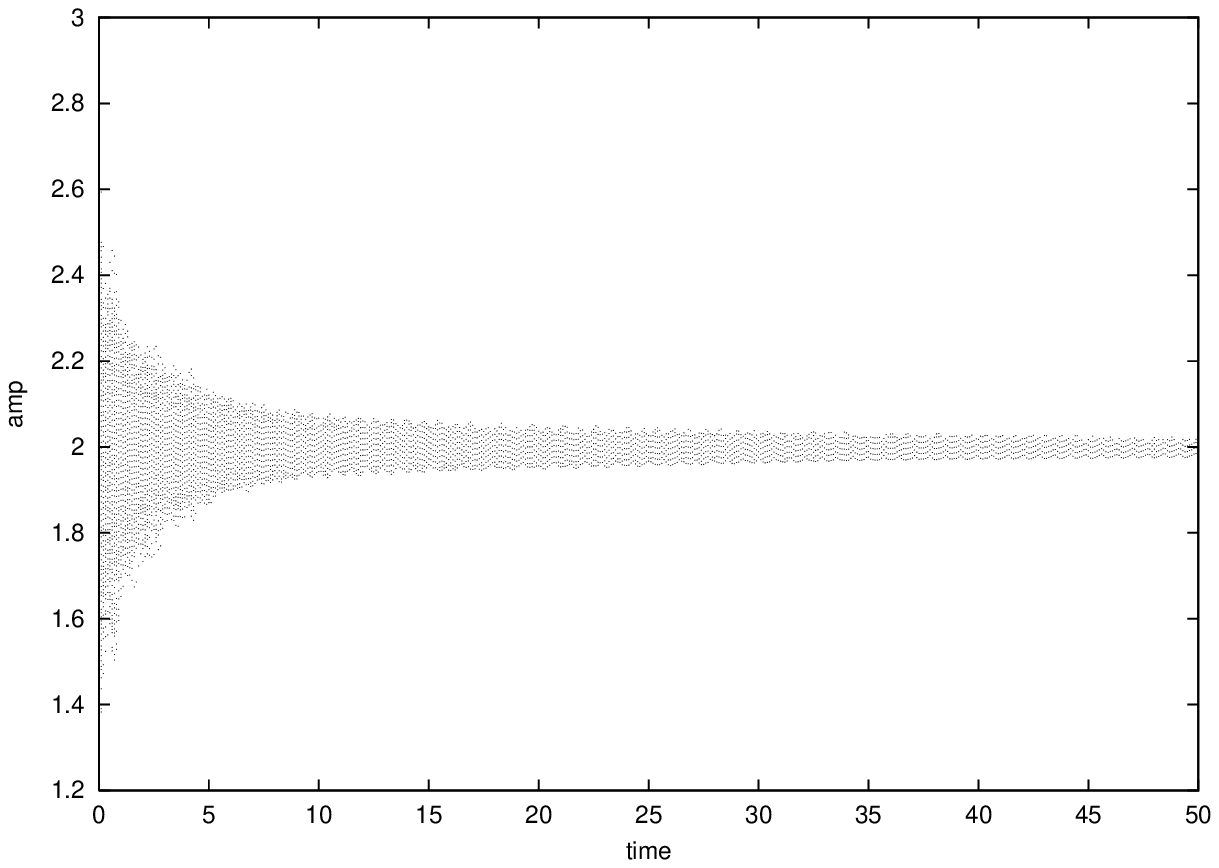}
       \end{center}
     \end{minipage}
   \end{tabular}
   \caption{ A target pattern of the numerical solution at time $t=50$ for the 
     initial condition given in Fig.\ref{fig:7} and $|W_n| ^2$ vs. $t$ for all
     $n$, which shows stability of the target pattern.  }
  \label{fig:8}
\end{figure}
 
\begin{figure}[htbp]
   \begin{tabular}{cc}
        \begin{minipage}{0.5\hsize}
       \begin{center}
         \includegraphics[width=5.0cm]{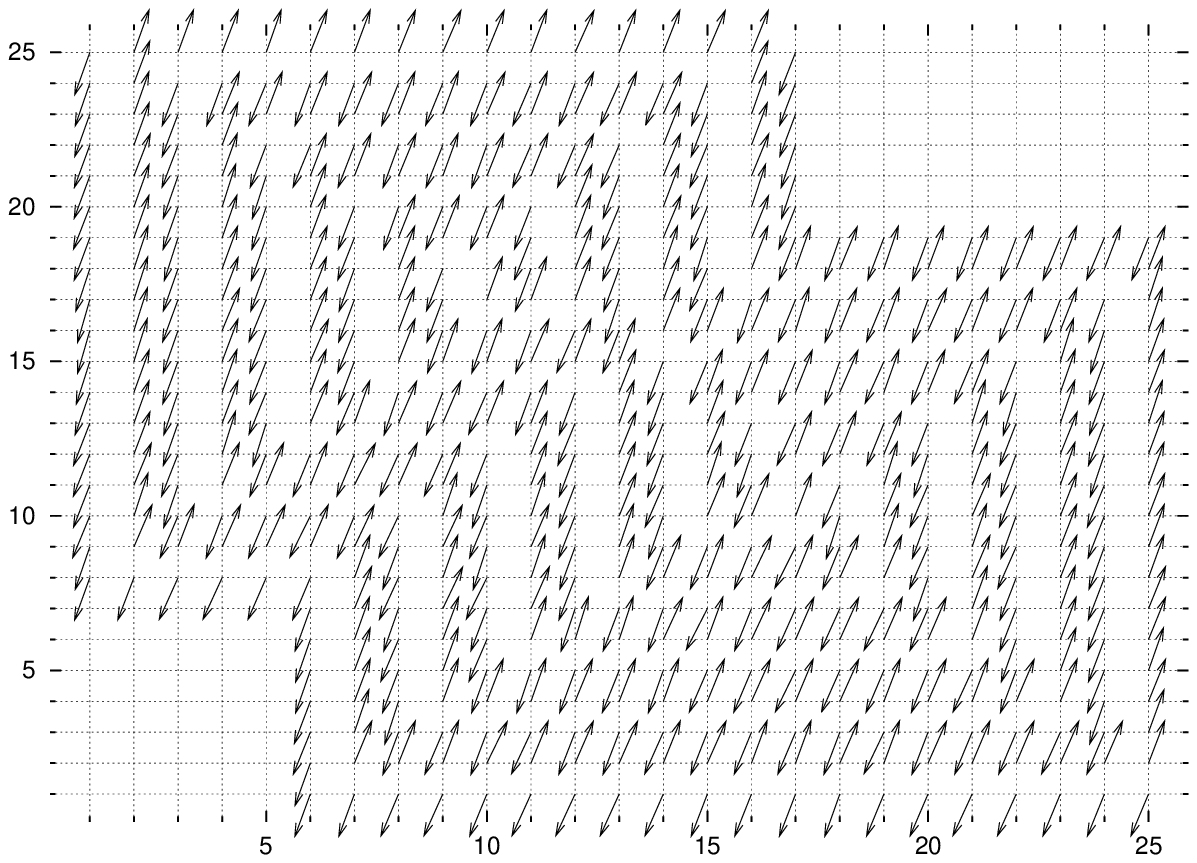}
       \end{center}
     \end{minipage}
     \begin{minipage}{0.5\hsize}
       \begin{center}
         \psfrag{amp}[][][0.6]{$|W|^2$}
         \psfrag{time}[][][0.6]{time}
         \includegraphics[width=5.0cm]{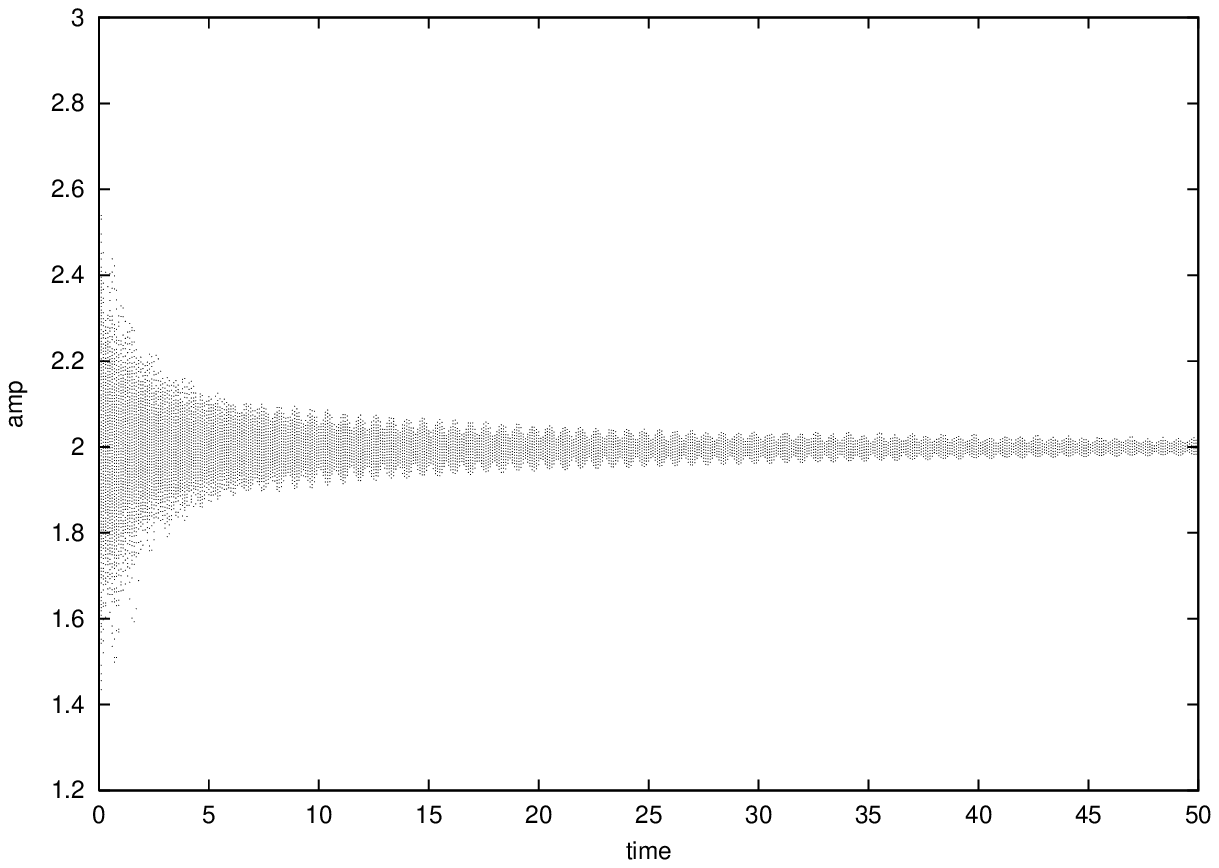}
       \end{center}
     \end{minipage}
   \end{tabular}
   \caption{A combined spiral pattern of the numerical solution at time $t=50$ for the 
     initial condition given in Fig.\ref{fig:7} and $|W_n| ^2$ vs. $t$ for all
     $n$, which shows stability of the combined spiral pattern.
   }
   \label{fig:9}
\end{figure}

\begin{figure}[htbp]
   \begin{tabular}{cc}
     \begin{minipage}{0.5\hsize}
       \begin{center}
         \includegraphics[width=5.0cm]{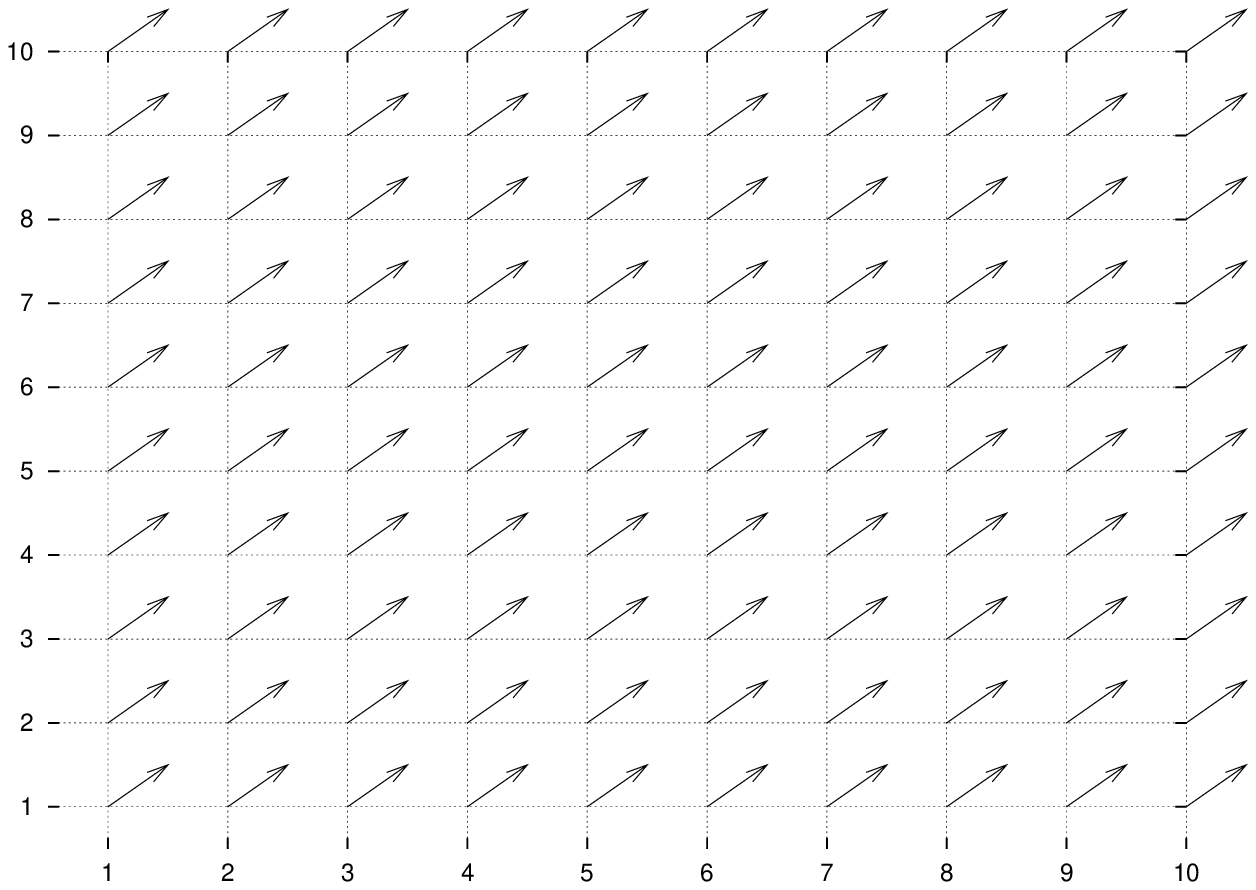}
       \end{center}
     \end{minipage}
     \begin{minipage}{0.5\hsize}
       \begin{center}
         \includegraphics[width=5.0cm]{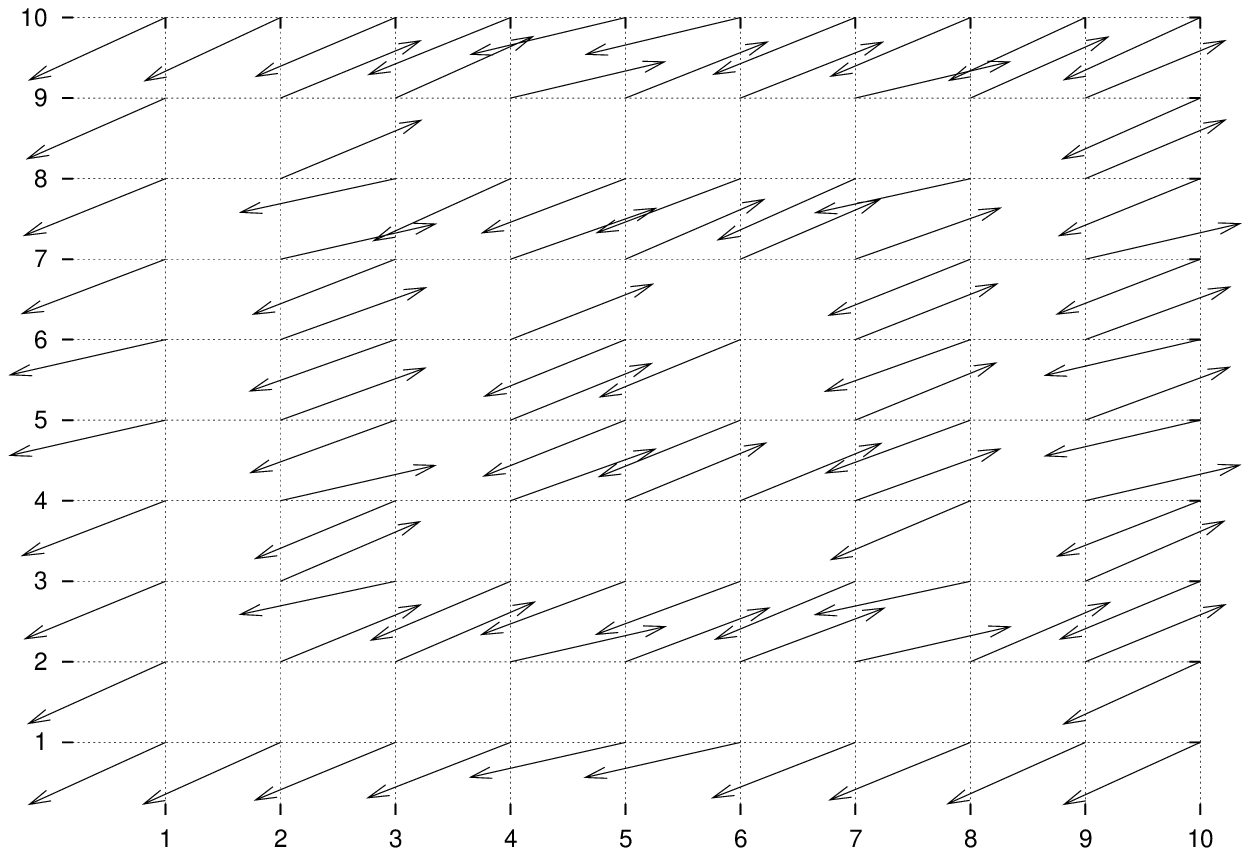}
       \end{center}
     \end{minipage}
   \end{tabular}
   \caption{An initial condition  $|W|=0.5$
     and  phase$= \pi /4$ for all oscillators and
     a target pattern at time $t=150$ developed from
     the initial condition.
       }
   \label{fig:13}
\end{figure}

Next, we show that a single target pattern can be reached
from some initial conditions far from the exact solution.
For example, setting initial values as  $W_n=1/2,\ \phi(n)=\pi/4$ for 
all $n$ in the case $p_1 =0.5$, $p_2 =3.0$, $p_3 =0.25$ and  $p_4 
=-0.5$, we eventually have a target pattern, where $|W_n|=2$, shown in  
Fig.\ref{fig:13}. We also find
  other initial conditions far from the exact solution that lead to
the single target pattern  but  those do not give enough informations 
for a basin structure of the target patterns.

%%%%%%%%%
\section{Van der Pol's Oscillators}
%%%%%%%%%%
Similar exact target and/or spiral solutions can be also constructed for
other diffusively and dispersively coupled systems of oscillators
which undergo the  subcritical Hopf bifurcation.
For example, we consider a system of van der Pol's oscillators
which are coupled with each other diffusively and dispersively on a
plane lattice.
\begin{eqnarray}
  \frac{d}{dt}x_n  &=&y_n+ax_n- \frac{a}{3}x_n^3+
  D_x \sum^4_{j \sim n}(x_j-x_n) \nonumber \\
  \frac{d}{dt}y_n  &=&-x_n+
  D_y \sum^4_{j \sim n}(x_j-x_n), \label{van-der}
   \end{eqnarray}
where $a$ is a positive constant and $D_x, D_y$ are diffusive
and dispersive constants respectively.
Let us seek  phase solutions of Eq.(\ref {van-der})
similar to those discussed earlier.
Assuming
\begin{equation}
  \sum^4_{j \sim n}x_j =0, \label{van-rel}
  \end{equation}
Eqs.(\ref {van-der}) are reduced to a single  van der Pol's equation
\begin{equation}
\frac{d^2}{dt^2}X_n+(X_n^2-1)\frac{d}{dt}X_n+X_n=0, \label{svan-der}
  \end{equation}
where $\sqrt{1+4D_y}t \rightarrow  t $,   $x_n=(1-4D_x/a)^{1/2} X_n$
and
\begin{equation}
(1+4D_y)>0, \quad (1-4D_x/a)>0 \label{van-cond}
\end{equation}
are assumed.
If the van der Pol's attractor of Eq.(\ref {svan-der}) is written as
$x_n(t)=A(t)$, its opposite-phase function $x_n(t)=-A(t)$ is also
the attractor of Eq.(\ref {svan-der}).
By choosing  $x_n(t)$ as $A(t)$ or $-A(t)$ for all $n$ so that
  $x_n(t)$ are satisfy the relation  (\ref {van-rel}),
  we can construct various exact solutions whose phase patterns
  are exactly the same as those in the discrete CGL equation.
  However, the stability analysis will be much more difficult than
  the previous case.
%%%%%%%%%
\section{Conclusions}
%%%%%%%%%%
We present various exact solutions of a discrete CGL equation on a 
plane lattice, which describe target patterns 
and spiral patterns and derive their stability criteria.
Exact solutions obtained here represent not only a single target pattern and  a 
single spiral pattern but also patterns consisting  
of targets and spirals. In the center of a spiral pattern, at least one 
defect is necessary and  a number of spiral's arms depends on a 
number of defects. For example, a defect produces a two-arm spiral 
and four defects in the center generate a four-arm spiral and so on.
For a system of the van der Pol's oscillators, we can also 
construct various exact solutions whose phase patterns
  are exactly the same as those in the discrete CGL equation.
pp
%%%%%%%%%%%%%%%%%%%%%%%%%%
\section* {Acknowledgement}
%%%%%%%%%%%%%%%%%%%%%%%%%%
One of the authors (K.N.) has been, in part, supported by a 
Grant-in-Aid for
Scientific Research
(C) 13640402 from the Japan Society for the Promotion of Science.
%%%%%%%%%%%%%%%%%%%%%%%%%%%%%%%%%%%%%%%%%%%%%%%%%%%%%%%%%%%%%%%%

%%%%%%%%%%%%%%%%%%%%%%%%%%%

%%%%%%%%%%%%%%%%%%%%%%

\end{document}